\documentclass[11pt,a4paper]{article}
\usepackage[hyperref]{acl2020}
\usepackage{times}
\usepackage{latexsym}
\usepackage{amsmath}
\usepackage{graphicx}
\usepackage{subcaption}
\usepackage[export]{adjustbox}

\usepackage{xcolor,xspace}

\newcommand{\norm}[1]{\left\lVert #1 \right\rVert}
\usepackage{mathtools}

\usepackage{microtype}

\aclfinalcopy 
\def\aclpaperid{2855} 

\newcommand{\structparagraph}[1]{}

\title{A Multi-Perspective Architecture for Semantic Code Search}

\author{Rajarshi Haldar$^{\dag}$, Lingfei Wu$^{\ddag}$, Jinjun Xiong$^{\ddag}$, Julia Hockenmaier$^{\dag}$ \\
$\dag$University of Illinois at Urbana-Champaign, Champaign, IL, USA\\ 
$\ddag$IBM Thomas J. Watson Research Center, Yorktown Heights, NY, USA\\
\{rhaldar2, juliahmr\}@illinois.edu, \{wuli, jinjun\}@us.ibm.com
}

\date{}

\begin{document}
\maketitle
\begin{abstract}
The ability to match pieces of code to their corresponding natural language descriptions and vice versa is fundamental for natural language search interfaces to software repositories. In this paper, we propose a novel multi-perspective cross-lingual neural framework for code--text matching, inspired in part by a previous model for monolingual text-to-text matching, to capture both global and local similarities. 
Our experiments on the CoNaLa dataset show that our proposed model yields better performance on this cross-lingual text-to-code matching task than previous approaches that map code and text to a single joint embedding space.
\end{abstract}



\section{Introduction}
\label{sec:introduction}

In semantic code search or retrieval, the user provides a natural language query, and the system returns a ranked list of relevant code snippets from a database or repository for that query. 
This task is usually performed using a matching model that computes the similarity between code snippets and natural language descriptions by mapping code and natural language embeddings into a common space where the distance between a piece of code and its corresponding description is small
\cite{gu2018deepcs, coacor}.

But current models do not explicitly model any interactions between the code and the description until the final step when their global similarity is calculated.

In this paper, we propose a novel multi-perspective neural framework for code--text matching that captures both global and local similarities. We show that it yields improved results on semantic code search. 

We apply our model to the CoNaLa benchmark dataset \cite{yin2018mining}, which consists of Python code snippets and their corresponding annotations in English. We believe that our model could be applied to other programming languages as well. We have made our code publicly available for research purpose 
\footnote{
\url{https://github.com/rajarshihaldar/codetextmatch}}.


\section{Background}
\label{sec:background}
Semantic code search is a cross-modal ranking problem where items in one modality (code) need to be ranked according to how well they match queries in another (natural language). 
One standard way to compute the similarity of items drawn from two different modalities or languages is to map each modality into a common ``semantic'' vector space such that matching pairs are mapped to vectors that are close to each other. 

\citet{gu2018deepcs} propose a code retrieval framework that jointly embeds code snippets and NL descriptions into a high dimensional embedding space such that the vectors representing a code snippet and its corresponding description have high similarity.

A variety of different approaches for learning embeddings for code have been proposed. Because source code is less ambiguous than natural language, there are ways to exploit the underlying structure of code to obtain better representations. \citet{Wan2019MultiModalAN,leclair2020improved} show that using features extracted from Abstract Syntax Trees (AST's) and Control Flow Graphs (CFG's) lead to creating better representations of code. \citet{Hu:2018:DCC:3196321.3196334,haque2020improved} show that ASTs represented as compact strings can be used to represent code. Following these approaches, we developed a multi-modal framework that generates embeddings for code using both the code tokens and an AST representation.


\section{Models}
\label{sec:code--text-model}
We compare four  models: a baseline model (\textbf{CT}) that only considers text and source code, a (\textbf{CAT}) model that also includes embedding of  Abstract Syntax Trees, a multi-perspective model (\textbf{MP}) that leverages multi-perspective matching operations as defined in a bilateral multi-perspective model \cite{DBLP:journals/corr/WangHF17}, and our  \textbf{MP-CAT} model that combines both \textbf{MP} and  \textbf{CAT} architectures. 

\subsection{CT: A Baseline Code and Text Model}
\label{sec:our-baseline-model}
Our baseline model (\textbf{CT}) is based on \citet{gu2018deepcs}'s CODEnn model. It maps both code and natural language descriptions to vectors in the same embedding space and then computes the similarity between these vectors using the L2 distance metric. 
These vectors are computed by two sets of three layers (one set per modality):

The \textbf{Word Embedding Module} consists of two independently pre-trained lookup tables that map code tokens or natural language tokens to embeddings. 
We use FastText  \cite{bojanowski-etal-2017-enriching}) for all embeddings in this paper. 

The \textbf{Context Representation Module} consists of bi-directional LSTM layers (one for code,  one for text) that map the word embedding sequences into another pair of sequences of embeddings that contain contextual information.

The \textbf{Maxpool Layer} performs max pool (separately per dimension) over the Context Representation embedding sequences to obtain a single vector.

The \textbf{Similarity Module} computes the similarity of the two vectors $v_c$ and $v_c$ produced by the Maxpool Layers as
\begin{align*}
    d(v_1, v_2) &= \sum_{i=1}^d(v_{1i}-v_{2i})^2\\
    \mathit{sim}(v_c,v_d) &=1-d(\frac{v_c}{\norm{v_c}_2}, \frac{v_d}{\norm{v_d}_2})
\end{align*}
\noindent
where $d$ returns the L2 distance between d-dimensional vectors $v_c$ and $v_d$.

\subsection{CAT: An AST-Based Model}
\label{sec:modeling-asts}
To capture both syntactic and semantic features, we augment our baseline \textbf{CT} model with embeddings based on the Abstract Syntax Tree (AST) representation of the code. Most programming languages, including Python, come with a deterministic parser that outputs the AST representation of a code snippet. Python has a library module called \href{https://docs.python.org/3/library/ast.html}{ast} that generates AST representations of code. We convert this AST representation to a string using structure-based traversal (SBT) \cite{Hu:2018:DCC:3196321.3196334}.
The \textbf{CAT} model is similar to the \textbf{CT} model, except that it extracts features from both the source code tokens and its corresponding AST representation. So the \textbf{Word Embedding Module} now contains three lookup tables: for code, AST, and natural language, respectively. Similarly, the \textbf{Context Representation Module} has 3 bi-directional LSTM layers which is followed by 3 \textbf{Maxpool Layers}. Before the output is passed to the similarity module, the output vectors of the two max pool layers representing code and AST are concatenated to form a single representation of the source code. Because of this, the hidden dimension in the bidirectional LSTM's of the \textbf{Context Representation Module} for the natural language sequence is double that of
code and AST sequences' LSTM hidden dimensions. This ensures that, after concatenation, the vectors representing the candidate code snippet and the natural language description are of the same dimension. After that, the \textbf{Similarity Module} computes the similarity of these vectors via the same L2-distance-based operation as in \textbf{CT}.

\subsection{MP: A Multi-Perspective Model}
\label{sec:bimpm} 
The \textbf{CT} and \textbf{CAT} models learn to map source code and natural language tokens into a joint embedding space such that semantically similar code-natural language pairs are projected to vectors that are close to each other. However, these two representations interact only in the final step when the global similarity of the sequence embeddings is calculated, but not during the first step when each sequence is encoded into its corresponding embedding. 
\citet{DBLP:journals/corr/WangHF17} show that, for tasks such as paraphrase identification and natural language inference that require two pieces of texts from the same language to compare, it is beneficial to include a number of different (i.e., multi-perspective) local matching operations between the two input sequences when computing their vector representations. 
Given contextual sequence encodings $P$ and $Q$ (computed, e.g., by biLSTMs) for the two sequences to be compared, \citet{DBLP:journals/corr/WangHF17}'s Bilateral Multi-Perspective Matching (BiMPM) model includes a matching mechanism that compares $P$ and $Q$ by matching each position in $P$ with all positions in $Q$, and by matching each position in $Q$ with all positions in $P$, under four different matching strategies. We will discuss these strategies in more detail under the Bilateral Multi-Perspective Matching (BiMPM) Module.

We apply the MP model to our cross-modal code-text matching task as follows: 
The \textbf{Word Embedding Layer} takes as input the code sequence, AST sequence, and description sequence. The output of this layer is three independent sequences of token embeddings, one for each input sequence.

The \textbf{Context Representation Module} consists of three sets of BiLSTM layers that each computes a contextual representation of each token in the corresponding input sequence. We concatenate the hidden states of the sequences representing the code and AST, respectively, to get one set of sequence embeddings representing the source code input.

The \textbf{Bilateral Multi-Perspective Matching (BiMPM) Module} compares the two sequences, say $P$ and $Q$, by matching each position in $P$ with all positions in $Q$, and by matching each position in $Q$ with all positions in $P$, under four different matching strategies $m$ that each produce new embedding sequences $P'_m$ and $Q'_m$ that have the same length as the original $P$ and $Q$. Each matching strategy is parameterized by a feedforward network (e.g.  $P'[i]_m = f^{P\rightarrow Q}_{m}(P[i], Q_m ; W_m^{P\rightarrow Q})$) that takes in a token embedding $P[i]$ and a strategy-specific single-vector representation of $Q_m$, and returns a new vector $P'[i]_m$ for $P[i]$. For each token $P[i] \in P$ (and conversely for any  $Q[j] \in Q$), $Q_m$ ($P_m$) is defined as follows:

\textbf{Full matching} sets $Q_m$ ($P_m$) to be the final hidden state of $Q$ (and vice versa for $P$).

\textbf{Maxpool matching} obtains $Q_m$ by performing maximum pooling (per dimension) across the elements of $Q$. 

\textbf{Attentive matching} computes $Q_m$ as a weighted average of all $Q[j] \in Q$, where $Q[j]$'s weight is the  
cosine similarity of $P[i]$ and  $Q[j]$. 

\textbf{Max-Attentive matching} sets $Q_m$ to be the $Q[j]$ with the highest cosine similarity to $P[i]$.

We concatenate the four $P'[i]_{m}$ ($Q'[i]_m$) for each token $i$ to get two new sequences $P'$ and $Q'$. 

The \textbf{Local Aggregation Module} aggregates these sequence embeddings into two fixed-length multi-perspective hidden representations by passing them through two different bi-LSTM layers (one for each sequence). For each sequence, we concatenate the final hidden states of both the forward and reverse directions to get a vector representation of that sequence.

The \textbf{Similarity Module} computes the similarity of the two vectors returned by the Aggregation Module as before. 

\subsection{MP-CAT: A Combined Model}

Our final model combines the MP and the CAT models. It contains the following components:

The \textbf{CAT} module reads in the code sequence, the AST sequence, and the natural language sequence and outputs two vectors, one jointly representing the code and the AST and the other representing the natural language description.

The \textbf{MP} module also reads in the code sequence, the AST sequence, and the natural language sequence. It returns two vectors, one for code and AST, and the other for the natural language description. The difference between this module and the previous is that \textbf{MP} contains local information that is ignored in the global \textbf{CAT} embeddings.

\begin{figure*}[t]
\centering
  \includegraphics[width=0.7\linewidth]{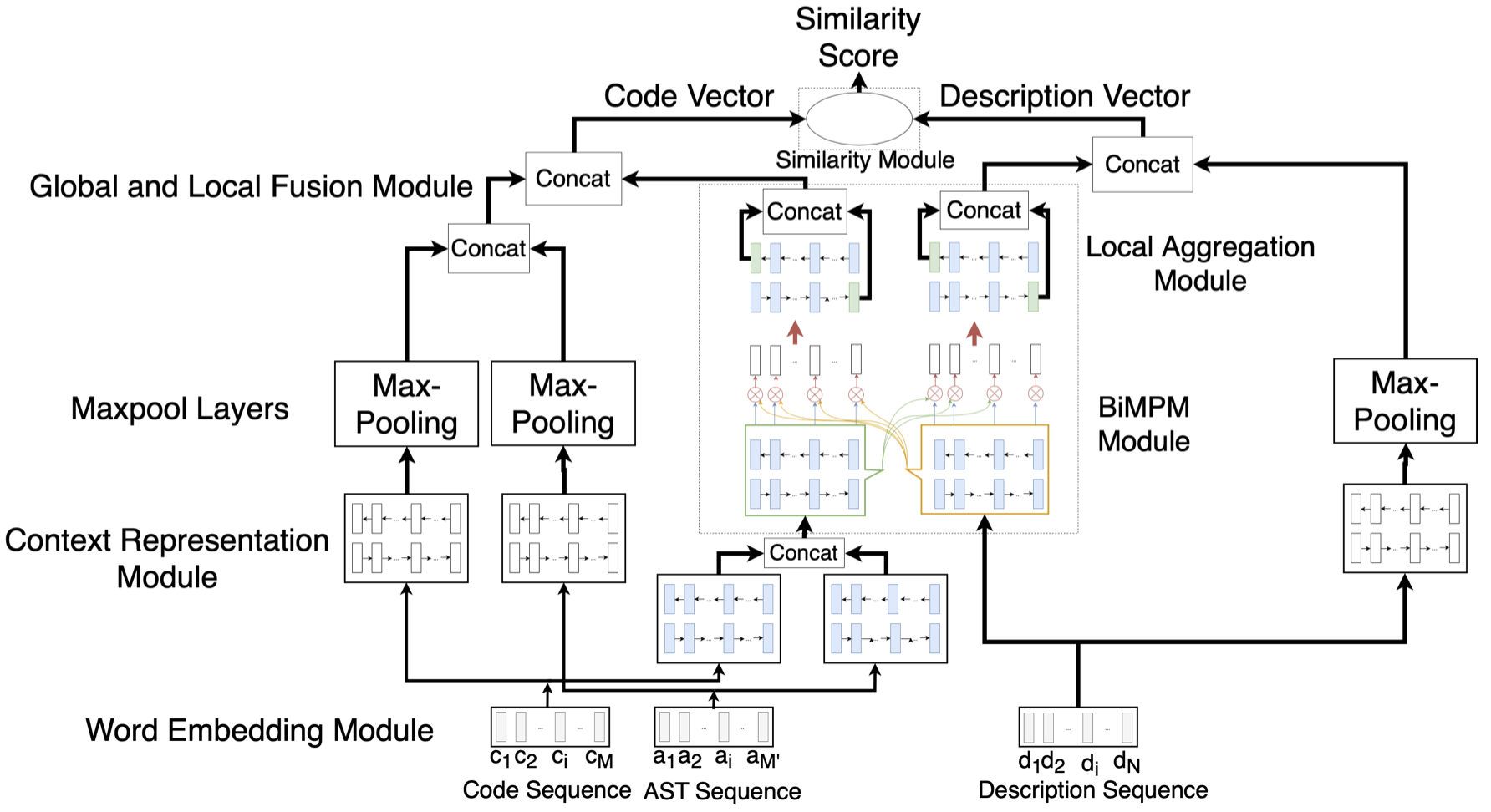}
  \caption{The MP-CAT framework that contains both global-level and local-level features for code--text matching}
  \label{fig:MPCTM}
\end{figure*}

The \textbf{Global and Local Fusion Module} concatenates the two CAT and MP vectors representing the code to get the final code representation, and does the same for the CAT and MP vectors representing the natural language description, before computing their L2 distance in the same manner as the other similarity modules. Figure ~\ref{fig:MPCTM} shows the pipeline of the MP-CAT framework.

\section{Experiments}
\label{sec:experiments}

\paragraph{The CoNaLa Dataset}
The CoNaLa dataset \cite{yin2018mining} has two parts, a manually curated parallel corpus of 2,379 training and 500 test examples, and a large automatically-mined dataset with 600k examples (which we ignore here). Each example consists of a snippet of Python code and its corresponding English description. 

\paragraph{Pre-processing}
\label{sec:preprocessing}
We pre-process the text representing both the source code and the natural language descriptions using sub-word regularization based on unigram language modeling \cite{kudo-2018-subword} transforms the original tokens into sequences of shorter (and hence more common) substrings. We use the sentencepiece library \cite{kudo-richardson-2018-sentencepiece} and follow the same approach as used by \citet{yin2018mining} for the CoNaLa dataset.

\paragraph{Training procedure}
During training, we use triplets consisting of a code snippet, a correct description, and an incorrect description (obtained by random sampling from the training set). We sample 5 incorrect descriptions for each code--text pair, giving us five triplets for each training example.
During the evaluation phase, for every natural language query $\mathcal{D}$, we calculate the rank of its corresponding code snippet $\mathcal{C}$ among all 500 candidates in the test set.

\begin{table}[]
\resizebox{0.5\textwidth}{!}{%
\begin{tabular}{|l|l|l|}
\hline
\textbf{Framework} & \textbf{Training Time (s)} & \textbf{Evaluation Time (s)} \\ \hline
CT                 & 4663.10                    & 6755.62                      \\ \hline
CAT                & 6702.69                    & 11050.68                     \\ \hline
MP                 & 183393.47                  & 17374.14                     \\ \hline
MP-CAT             & 240062.38                  & 25306.97                     \\ \hline
\end{tabular}%
}
\caption{Training and Evaluation times for all our models. The models were trained for 100 epochs and the evaluation time was computed on 500 test queries.}
\label{tab:running-time}
\end{table}

\begin{table}[]
\resizebox{0.5\textwidth}{!}{%
\begin{tabular}{|l|l|l|l|l|}
\hline
\textbf{Frameworks} & \textbf{MRR}   & \textbf{R@1} & \textbf{R@5} & \textbf{R@10} \\ \hline
CT                  & 0.172          & 7.4               & 24.0              & 39.6               \\ \hline
CAT                 & 0.207          & 9.0               & \textbf{32.2}     & 45.0               \\ \hline
MP                  & 0.154          & 6.4               & 21.6              & 33.6               \\ \hline
MP-CAT              & \textbf{0.220} & \textbf{11.0}     & \textbf{32.2}     & \textbf{47.4}      \\ \hline
\end{tabular}%
}
\caption{Code Search Results}
\label{tab:retrieval-results}
\end{table}

\subsection{Experimental Setup}
\label{sec:experimental-setup}
We train our  models on triplets $\langle C, D^+, D^- \rangle$ consisting of a snippet of code $C$, a natural language description $D^+$ that correctly describes what the code does (a positive example), and a description $D^-$ that does not describe what the code does (a negative example).  We minimize the ranking loss with margin $\epsilon$, following \citet{gu2018deepcs}:

\begin{small}
\begin{equation*}
    \mathcal{L}(\theta)=\!\!\!\sum_{\langle C,D^{+},D^{-}\rangle\!}\!\!\!\max \big(0, \epsilon-\cos (C, D^+)+\cos (C, D^-)\big)
\end{equation*}
\end{small}
In the CAT model, since we  first concatenate the vectors for the code and AST before comparing them with the vector for the natural language description, the first two vectors are each half the dimension size of the third one. Our models are implemented in PyTorch \cite{paszke2017automatic} and  trained using  Adam  \cite{Kingma2015AdamAM}. 

Each model is trained for 100 epochs, and during the evaluation step, we use a set of 500 natural language queries from the test set. The training and evaluation times are shown in Table~\ref{tab:retrieval-results}.

\subsection{Results}
\label{sec:results}
Table~\ref{tab:retrieval-results} shows our test set results for code search. We report Recall@K (K=1,5,10) and mean reciprocal rank (MRR) of the correct answer.

\paragraph{The Impact of Modeling ASTs:}
In going from the first (CT) row to the second (CAT) row in Table~\ref{tab:retrieval-results}, we see that the AST features alone increase MRR from 0.172 to 0.207. There is also an increase in R@k for all values of k. In fact, its R@5 values are competitive with our best model.

\paragraph{Multi-Perspective Results:}
\label{sec:bimpmresults}

The results for the multi-perspective models are both surprising and interesting. Row 3 of Table~\ref{tab:retrieval-results} shows that the MP model on its own under-performs and actually has the worst results out of all the models we tested. On the other hand, we see that combining the MP and the CAT models into one framework gives the best performance across the board. This shows that even if we use a multi-perspective framework to model local features, we still need encoders to capture the global features of code and text in addition to the local features; otherwise, we end up missing the forest for the trees.

\begin{table}[]
\resizebox{0.5\textwidth}{!}{%
\begin{tabular}{|l|l|l|}
\hline
\textbf{Query}                                                                                                                                  & \textbf{MP-CAT}                                                                                       & \textbf{CAT}                                                                                         \\ \hline
\begin{tabular}[c]{@{}l@{}}Sort dictionary `x` by value in\\ ascending order\end{tabular}                                                       & \begin{tabular}[c]{@{}l@{}}sorted(list(x.items( )),\\ key = operator.itemgetter(1))\end{tabular}      & \begin{tabular}[c]{@{}l@{}}for k in sorted(\\ foo.keys( )):\\     pass\end{tabular}                     \\ \hline
\begin{tabular}[c]{@{}l@{}}Run a command `echo hello world`\\ in bash instead of shell\end{tabular}                                           & \begin{tabular}[c]{@{}l@{}}os.system\\  (/bin/bash -c "echo hello world")\end{tabular}               & \begin{tabular}[c]{@{}l@{}}os.system\\ ( 'GREPDB=\\"echo 123";\\ /bin/bash -c "\$GREPDB"')\end{tabular}   \\ \hline
\begin{tabular}[c]{@{}l@{}}Select records of dataframe\\ `df` where the sum of column\\ 'X' for each value in column 'User'\\ is 0\end{tabular} & \begin{tabular}[c]{@{}l@{}}df.groupby('User'){[}'X'{]}.filter(\\ lambda x: x.sum() == 0)\end{tabular} & \begin{tabular}[c]{@{}l@{}}print(df.loc{[}df{[}'B'{]}.isin(\\ {[}'one', 'three'{]}){]})\end{tabular} \\ \hline
\end{tabular}%
}
\caption{The top hits returned by the MP-CAT and CAT models for a natural language query.}
\label{tab:mpcat-vs-cat}
\end{table}

\begin{table}[]
\resizebox{0.5\textwidth}{!}{%
\begin{tabular}{|l|l|l|}
\hline
\textbf{Query}                                                                                                                      & \textbf{MP-CAT}                                                                              & \textbf{MP}                                                                                                                                                                                \\ \hline
\begin{tabular}[c]{@{}l@{}}Concatenate elements of a\\ list 'x' of multiple integers\\ to a single integer\end{tabular}             & \begin{tabular}[c]{@{}l@{}}sum(d*10**i\\ for i, d in enumerate(\\ x{[}::-1{]}))\end{tabular} & \begin{tabular}[c]{@{}l@{}}{[}float( i )\\ for i in lst{]}\end{tabular}                                                                                                                    \\ \hline
\begin{tabular}[c]{@{}l@{}}convert pandas DataFrame\\ `df` to a dictionary using\\ `id` field as the key\end{tabular}               & \begin{tabular}[c]{@{}l@{}}df.set\_index( 'id').to\_dict()\end{tabular}                    & \begin{tabular}[c]{@{}l@{}}data{[}\\ data{[}'Value'{]} == True{]}\end{tabular}                                                                                                             \\ \hline
\begin{tabular}[c]{@{}l@{}}Replace repeated instances\\ of a character '*' with a\\ single instance in a string 'text'\end{tabular} & re.sub('\textbackslash{}\textbackslash{}*\textbackslash{}\textbackslash{}*+', '*', text)     & \begin{tabular}[c]{@{}l@{}}re.sub('\textasciicircum{}((\\ ?:(?!cat).)*cat(\\ ?:(?!cat).)*)cat',\\ '\textbackslash{}\textbackslash{}\textbackslash{}\textbackslash{}1Bull', s)\end{tabular} \\ \hline
\end{tabular}%
}
\caption{The top hits returned by the MP-CAT and MP models for a natural language query.}
\label{tab:mpcat-vs-mp}
\end{table}

\paragraph{Comparison of MP-CAT, MP and CAT Models}
In Table~\ref{tab:mpcat-vs-cat}, we present the retrieval results for select natural language queries from the development set returned by the MP-CAT and CAT models. We do the same thing for MP-CAT and MP models in Table~\ref{tab:mpcat-vs-mp}. Comparing MP-CAT and CAT, we observe that while CAT correctly identifies the data structures and libraries required to solve the user's problem, it ends up returning the wrong command. MP, on the other hand, sometimes fails to identify even the correct libraries required. In the second example in Table~\ref{tab:mpcat-vs-mp}, it fails to understand that there is also a dictionary involved and ends up returning the wrong command. MP-CAT successfully finds the required code snippet when the user queries are longer and have multiple data structures involved.


\section{Conclusions}
\label{sec:conclusions}
In this paper, we consider the task of semantic code search or retrieval using a code--text similarity model. We propose MP-CAT, a novel multi-perspective deep neural network framework for this task. In contrast to previous approaches, the multi-perspective nature of our model allows it to capture richer similarities between the two sequences.

\section*{Acknowledgement}
This work is supported by the IBM-ILLINOIS Center for Cognitive Computing Systems Research (C3SR), a research collaboration as part of the IBM AI Horizons Network.
\bibliography{acl2020}
\bibliographystyle{acl_natbib}

\end{document}